# Full Crystallographic Imaging of Hexagonal Boron Nitride Monolayers with Phonon-Enhanced Sum-Frequency Microscopy

Niclas S. Mueller[a†], Alexander P. Fellows[a†], Ben John[a], Andrew E. Naclerio[b], Christian Carbogno[c], Katayoun Gharagozloo-Hubmann[d], Damián Baláž[c], Ryan A. Kowalski[e,f], Hendrik H. Heenen[c], Christoph Scheurer[c], Karsten Reuter[c], Joshua D. Caldwell[e,f], Martin Wolf[a], Piran R. Kidambi[b], Martin Thämer[a,*], Alexander Paarmann[a,*]

[a] Department of Physical Chemistry, Fritz-Haber-Institute of the Max-Planck-Society, 14195 Berlin, Germany

[b] Department of Chemical and Biomolecular Engineering, Vanderbilt University, Nashville, TN, 37235 USA

[c] Theory Department, Fritz-Haber-Institute of the Max-Planck-Society, 14195 Berlin, Germany

[d] Department of Physics, Freie Universität Berlin, 14195 Berlin, Germany

[e] Department of Mechanical Engineering, Vanderbilt University, Nashville, TN, 37235 USA

[f] Interdisciplinary Materials Science Program, Vanderbilt University, Nashville, TN, 37235 USA

[†] contributed equally

*corresponding authors (thaemer@fhi-berlin.mpg.de, alexander.paarmann@fhi-berlin.mpg.de)

**Hexagonal boron nitride (hBN) is an important 2D material for van der Waals heterostructures, single photon emitters, and infrared nanophotonics. The optical characterization of mono- and few-layer samples of hBN however remains a challenge as the material is almost invisible optically. Here we introduce phase-resolved sum-frequency microscopy as a technique for imaging monolayers of hBN grown by chemical vapor deposition (CVD) and visualize their crystal orientation. A combination of femtosecond mid-infrared (IR) and visible laser pulses is used for sum-frequency generation (SFG), which is imaged in a wide-field optical microscope. The IR laser resonantly excites a phonon of hBN that leads to an ~800-fold enhancement of the SFG intensity, making it possible to image large 100x100 µm² sample areas in less than 1 s. Implementing heterodyne detection in combination with azimuthal rotation of the sample further provides full crystallographic information. Through combined knowledge of topography and crystal orientation, we find that triangular domains of CVD-grown monolayer hBN have nitrogen-terminated zigzag edges. Overall, SFG microscopy can be used as an ultra-sensitive tool to image crystal structure, strain, stacking sequences, and twist angles, and is applicable to the wide range of van der Waals structures, where location and identification of monolayer regions and interfaces with broken inversion symmetry is of paramount importance.**

Hexagonal boron nitride (hBN) is one of the most widely used van der Waals (vdW) materials because of its pivotal role as a substrate, encapsulating material, and tunneling barrier in vdW heterostructures.[1,2] The material furthermore hosts single photon emitters that can be used as quantum light sources at room temperature.[3,4] Moreover, hBN is widely applied for mid-infrared (IR) nanophotonics because of hyperbolic phonon polaritons that enable nanoscale waveguiding and strong confinement of IR light.[3,5-7] Besides this pronounced mid-IR optical response, hBN is optically transparent across the entire near-IR and visible (VIS) spectral range due to its >5 eV bandgap. This makes it very hard to locate and characterize mono- and few-layers of this material with optical techniques.[8] The Raman response of an hBN monolayer, for example, is ~50 times weaker than that of monolayer graphene, while the more sensitive stimulated and coherent Raman scattering techniques lack crystallographic information for hBN.[8-10]

Fortunately, hBN can be also characterized with 2$^{nd}$-order nonlinear techniques, which are sensitive to its crystal structure. This is possible because monolayers and few-layers with odd layer numbers, as well as twisted interfaces, have broken inversion symmetry.[11-14] Second harmonic generation (SHG) has emerged as a key tool to characterize the crystal orientation and twist angles of many 2D vdW materials.[15-20] Due to the selection rules of SHG, a characterization of the polarization dependence gives access to the crystallographic axes.[11] This method has also been recently used to image the twist angles and stacking configurations of buried interfaces between hBN films.[13] There are, however, two challenges that prevent SHG from becoming a standard tool for hBN characterization: (1) The SHG intensity from hBN mono- and few-layers is about 1000 times weaker than that of transition metal dichalcogenide (TMD) monolayers.[11] This is because hBN is lacking optical resonances in the VIS and near-IR, while TMDs have excitonic resonances that enhance the SHG response. (2) An SHG intensity measurement does not provide full crystallographic information, i.e. it detects a 6-fold symmetry whereas the lattice of hBN monolayers has 3-fold symmetry, and therefore cannot distinguish B-N from N-B crystal directions.[11]

Here, we establish sum-frequency generation (SFG) microscopy for addressing these two challenges. We use our recently developed wide-field optical microscopy (Refs. 21,22) to image SFG signals generated from the combined illumination of the sample with fs laser pulses in the mid-IR and VIS spectral ranges. The mid-IR laser covers a phonon resonance of hBN monolayers with the bandwidth of the IR pulse, which leads to an ~800-fold intensity enhancement, making SFG equally efficient as SHG in excitonic TMDs. This enables rapid screening of substrates for hBN monolayers. Additionally, by heterodyning the SFG signal with a local oscillator and using balanced imaging,[21] we measure the amplitude and phase of the SFG signal with high sensitivity, giving access to the full crystal orientation. We demonstrate this by imaging hBN monolayer islands on a transparent fused silica substrate that were grown by chemical vapor deposition (CVD) on a catalytic substrate before transfer.[23,24] By analyzing the SFG signal when azimuthally rotating the sample, we image the topography and crystal orientation simultaneously and find that the hBN islands have nitrogen-terminated zigzag edges. The technique of SFG microscopy is applicable to the wide range of vdW heterostructures containing hBN and may be used to image strain, stacking order, and twist angles in the future.

## Results

### Phase-Resolved Sum-Frequency Microscopy

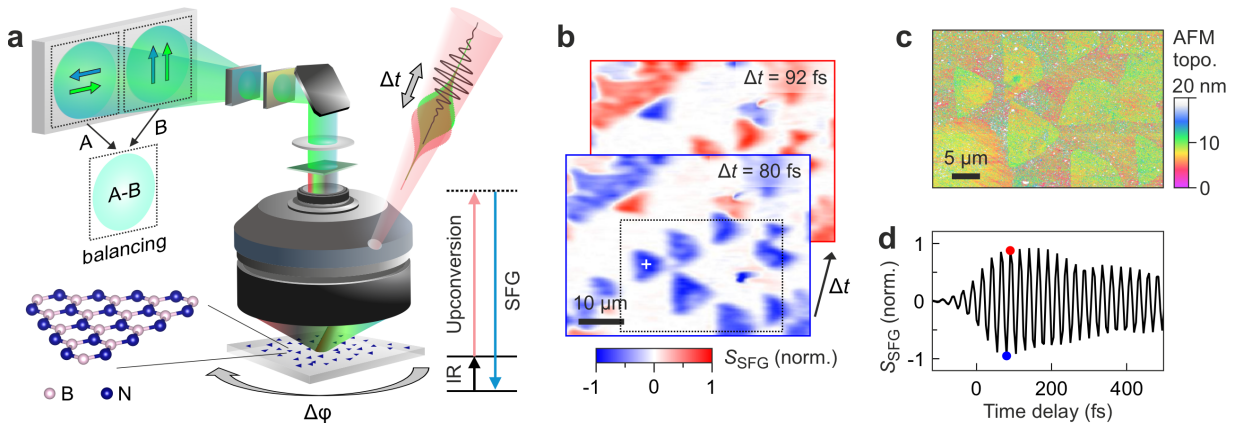

**Figure 1. Phase-Resolved SFG Microscopy of hBN.** (a) Schematic of the SFG microscope. An IR beam (black) and VIS upconversion beam (red) with controllable time delay $\Delta t$ collinearly illuminate a large sample area (150x200 µm) through a hole in a reflective objective. The nonlinear SFG signal (blue) is collected by the objective and guided through filters to a CCD camera. The SFG light is interfered with a collinear local oscillator (green) for varying time delays to obtain phase and spectral information. Two interference images with orthogonal polarization components (A and B) are recorded and balanced (A-B) to measure the SFG amplitude. The sample can be azimuthally rotated by $\Delta \varphi$ (SI Video 1). The inset shows a sketch of the hBN crystal structure. (b) SFG amplitude images of hBN monolayer flakes grown by CVD recorded for two time delays $\Delta t$, indicated by red and blue dots in (d). (c) AFM topography image of the area in (b) that is highlighted with a dotted box. (d) SFG amplitude as a function of IR-VIS time delay $\Delta t$, evaluated at a single pixel shown as white cross in (b).

The hBN monolayers are imaged and characterized with phase-resolved SFG microscopy (Fig. 1a, Methods, Refs. 21,22,25). An ultrafast laser is used to generate two broadband beams: one in the mid-IR (~7 µm, ~100 fs, ~5 µJ, *p*-polarization) and a second in the visible frequency range for upconversion (~690 nm, ~100 fs, ~3 µJ, *p*-polarization). The two beams are then combined and collinearly directed through a custom-made hole in a reflective Schwarzschild objective to illuminate the entire field-of-view.[21] The two laser pulses generate an SFG signal at the sample that is imaged with a CCD camera. In order to obtain phase information, a local oscillator (LO) beam that has the same frequency as the SFG signal is further guided collinearly into the microscope.[21,26,27] The reflected LO then interferes with the SFG signal from the sample, with the interference term being isolated through paired-pixel balanced imaging in order to extract a signal $S_{\text{SFG}}$ proportional to the SFG amplitude (Methods).[21] For this, two images are simultaneously generated on the CCD camera that have opposite signs of the interference cross term of SFG and LO. Subtracting the two images and referencing yields a balanced image of $S_{\text{SFG}}$ with strongly reduced noise that contains phase information (Fig. 1b).

We use the SFG microscope to characterize CVD-grown monolayers of hBN,[23,24,28] that were transferred to a fused silica substrate (Fig. 1b-d). On this transparent substrate the monolayers are optically undetectable in a standard linear reflection microscope (Fig. S4a) because of the ~3.3 Å thickness of the monolayers and no optical resonances of hBN in the visible spectral range.[8] In contrast, the SFG microscope enables imaging of hBN monolayers across a large (~200x150 µm$^2$) sample area in <1 s (Fig. 1b, Fig. S4b). The images clearly resolve the triangular shape of the hBN monolayer islands as the spatial resolution (~1 µm here) is limited by the visible SFG wavelength, which is much shorter than the mid-IR diffraction limit at $\lambda_{\text{IR}}$~7 µm.[21,22,29] A comparison to an AFM topography image of a selected smaller area (Fig. 1c, >30 min acquisition time) demonstrates the high chemical selectivity of SFG microscopy to the hBN monolayers, whereas the AFM image is also affected by contamination and surface roughness, which strongly reduces the contrast. The different signs and magnitudes of $S_{\text{SFG}}$ from the hBN islands already hint to different crystal orientations, as shown below.

To obtain spectral information, an interferogram is recorded for each image pixel by scanning a delay of the mid-IR pulse with respect to the VIS and LO pulses (Fig. 1b,d, SI Video 2).[26,30] As such, the technique can be understood as a pixel-wise, nonlinear-optical version of Fourier-transform IR (FTIR) spectroscopy, but additionally providing phase information. The SFG amplitude directly probes the IR field-resolved optical response of the material, which here is dominated by the ringing of the transverse optical IR phonon resonance of hBN (Fig. 1d). A similar response was also recently probed by phonon-enhanced four-wave mixing.[31] The spectral response can then be obtained from a Fourier transformation (see below). As this spectral response is probed simultaneously for the entire wide-field microscope image, the SFG microscope enables a direct correlation of spatial and spectral information through hyperspectral imaging.

**Phonon-Enhanced Sum-Frequency Generation**

The IR phonon resonance of hBN at ~7.3 µm leads to a strong enhancement of the SFG signal (Fig. 2). We measure an integrated SFG response of hBN monolayers that is ~10x larger than the off-resonant response of a $z$-cut α-quartz substrate that is used as a phase and amplitude reference (Fig. 2a). This is remarkable as bulk quartz has broken inversion symmetry leading to an effective SFG probing depth (coherence length) of ~36 nm which is ~100x thicker than the hBN monolayer (SI Section S4). After a Fourier transformation and referencing with quartz,[30] we obtain the spectrum of the effective nonlinear susceptibility of monolayer hBN, which shows a pronounced resonance from the $E'$ transverse optical (TO) phonon of hBN at $\omega_{TO}$ = 1368 cm$^{-1}$ (Fig. 2b). The spectral dependence is well explained by (see SI Section S1 for full tensorial expression)[32,33]

$$\chi^{(2)}(\omega_{IR}) = \chi^{(2)}_\infty \left(1 + \frac{\alpha_R Z^*}{2V_{uc}M} \frac{1}{\omega_{TO}^2 - \omega_{IR}^2 - i\gamma_{TO}\omega_{IR}}\right), \quad (1)$$

where $\chi^{(2)}_\infty$ is the off-resonant susceptibility from electronic transitions at larger frequencies, $\omega_{IR}$ the IR laser frequency, $\gamma_{TO}$ the damping of the phonon, $V_{uc}$ the unit cell volume, $M$ the effective mass, $\alpha_R$ the Raman polarizability, and $Z^*$ the Born effective charge (Fig. 2b, dotted lines). This shows that the phonon needs to be both Raman- and IR-active in order to enhance the SFG response, which is generally the case for materials with broken inversion symmetry, like hBN monolayers.[32-34] The phonon-enhanced $\chi^{(2)}$ is particularly large in hBN monolayers because of the strongly anharmonic potential associated with the atomic vibrations.[35] On resonance, we obtain an effective surface susceptibility of $\chi^{(2)}_{eff}(\omega_{TO}) = 1.9 \cdot 10^{-19}$ m$^2$/V, which corresponds to a bulk susceptibility of $\chi^{(2)}(\omega_{TO}) \approx 580$ pm/V when assuming a thickness of $d_{hBN}$ = 0.33 nm for monolayer hBN. This is much larger than the off-resonant susceptibility $\chi^{(2)}_\infty$ = 20.8 pm/V measured previously by SHG on hBN monolayers using near-IR lasers.[11] Comparing to this off-resonant response, we find a ~790-fold enhancement of the SFG intensity by the IR phonon resonance (Fig. 2c, SI Section S4). This makes SFG of hBN monolayers equally efficient as SHG of transition metal dichalcogenides, which have excitonic resonances in the visible spectral range, and where SHG emerged as a standard tool to characterize crystal orientations and twist angles.[15,16,18-20,36]

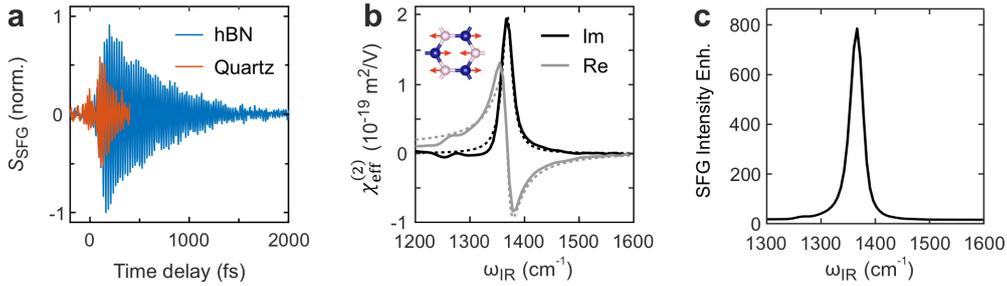

**Figure 2. Phonon-Enhancement of SFG.** (a) Time-domain interferogram of SFG amplitude for monolayer hBN (blue, evaluated at white cross in Fig. 1b) and *z*-cut α-quartz (orange) by scanning a time delay of the mid-IR pulse with respect to the VIS and LO pulses. (b) Measured SFG spectrum (solid lines) of monolayer hBN with the E' TO phonon resonance at $\omega_{TO}$ = 1368 cm$^{-1}$ (see inset). A fit with Eq. (1) is shown as dashed lines. (c) Enhancement of the SFG intensity by the IR phonon resonance of hBN. The resonant $|\chi^{(2)}_{eff}(\omega_{IR})|^2$ was referenced by the off-resonant $|\chi^{(2)}_{eff}|^2$ measured in Ref. 11.

## Full Crystallographic Imaging of hBN Monolayers

Beyond visualization of hBN monolayers, phase-resolved SFG microscopy gives access to their full crystallographic orientation (Fig. 3). When rotating the sample azimuthally with respect to the laser polarization, the resonant SFG amplitude at $\omega_{IR} = \omega_{TO}$ oscillates and follows the 3-fold symmetry of the $D_{3h}$ crystal lattice of monolayer hBN, with $S_{SFG,x} \propto \cos(3\varphi)$, where $\varphi$ is the angle between the arm-chair crystal direction and the in-plane component of the laser polarization (Fig. 3a, SI Section S1-3). The SFG signal is thus largest when the laser beams are polarized along the arm-chair crystal directions, and possesses positive sign along the B-N direction vs negative sign along N-B (Fig. 3a,b). In contrast to previous SHG intensity measurements which detected a 6-fold symmetry $|\cos(3\varphi)|^2$,[11-14] the heterodyned measurements here allow us to determine the full crystal orientation.[17]

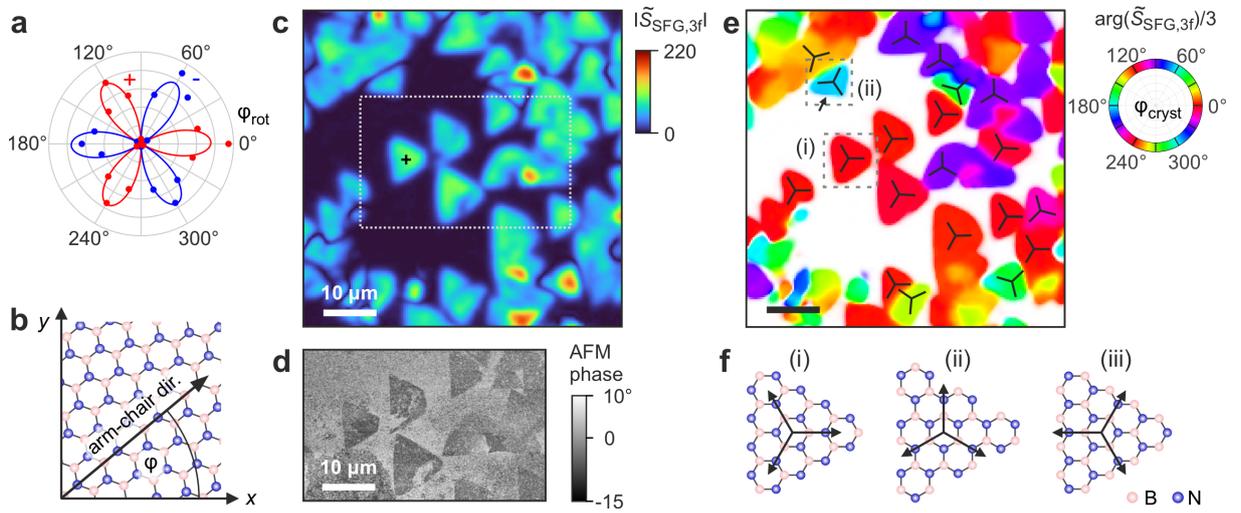

**Figure 3. Crystallographic Imaging of hBN Monolayers.** (a) Resonant SFG amplitude $\text{Im}[S_{SFG}(\omega_{IR} = \omega_{TO})]$ of monolayer hBN [cross in (c)] as function of azimuthal sample rotation angle $\varphi_{rot}$, with red indicating positive and blue negative sign. (b) Angle $\varphi = \varphi_{cryst} - \varphi_{rot}$ between crystallographic B-N arm-chair direction and in-plane polarization of all beams along *x*. (c) Image of SFG amplitude $|\tilde{S}_{SFG,3f}|$ with 3-fold rotational symmetry obtained from pixel-wise Fourier analysis of $\text{Im}[S_{SFG}(\varphi_{rot})]$ at $\omega_{IR}$ = 1368 cm$^{-1}$ (see Fig. S5 for full measurement area). (d) AFM phase image of area in (c) that is highlighted with gray dotted box. Compared to an AFM topography image of the same area in Fig. 1c the phase image is more sensitive to changes in composition and layer number. (e) Image of rotational phase $\arg(\tilde{S}_{SFG,3f})/3 = \varphi_{cryst}$ of the SFG amplitude with 3-fold rotational symmetry obtained with similar analysis as for (c). Black lines show the arm-chair crystal directions and colors the angle $\varphi_{cryst}$ between the arm-chair direction and *x* axis. (f) Orientation of B-N arm-chair direction (arrows) with respect to triangle edges depending on edge termination, with (i) N-terminated zigzag edges, (ii) arm-chair edges and (iii) B-terminated zigzag edges. Dashed boxes in (e) show flakes with (i) three N-terminated zigzag edges and (ii) two N-terminated zigzag and one arm-chair edge (see arrow), while B-terminated zigzag edges are not observed in our sample.

In order to visualize the topography and crystal orientation of all hBN islands, we use a rotational Fourier analysis of the SFG images recorded for different azimuthal sample rotations (Figs. 3c,e, Methods and SI).[22,25] For each image pixel, this approach extracts the experimental SFG signal component $\tilde{S}_{\text{SFG},3f}$ with 3-fold symmetry, and thus probes the $D_{3h}$ crystal symmetry of hBN. The Fourier analysis provides both a magnitude image $|\tilde{S}_{\text{SFG},3f}|$ that shows the topography (Fig. 3c) and a rotational phase image $\arg(\tilde{S}_{\text{SFG},3f})/3$ that shows the crystal orientation of all hBN islands (Fig. 3e). In contrast to a single sample rotation, $|\tilde{S}_{\text{SFG},3f}|$ visualizes all hBN islands independent of their individual crystal orientation (compare Figs. 1b and 3c). The SFG image thus provides detailed information about the shape and structure of the hBN islands, and correlates well with the topography probed in an AFM phase measurement of a selected smaller sample area, even resolving small morphological imperfections (compare Figs. 3c dashed box with 3d). Several small areas that appear as bright spots in the SFG amplitude most likely correspond to bi- and few-layers with nonzero twist angles, which leads to an increase of the SFG signal (SI Fig. S6).

The rotational phase image directly visualizes the local crystal orientation of the hBN monolayer islands, which were grown into a partially continuous film by CVD (Fig. 3e). Most of the individual islands are single-crystals and have triangular shapes, which enables a correlation of their shape with the crystal orientation. Analysis of these images shows that for almost all islands, the arm-chair crystal directions point towards the corners of the triangles, which implies that the islands have zigzag edges [Fig. 3f(i) and 3e black lines]. We also observe a few flakes with irregular shapes that have arm-chair edges, Fig. 3e,f(ii). The prevalence of zigzag edges was also suggested by CVD growth models which revealed that zigzag edges are energetically more favorable.[23,37,38] Phase-resolved SFG enables us to distinguish islands with opposite (180° rotated) crystal orientations, which would have identical signal in an intensity measurement. Through comparison with density functional theory calculations (SI Section S2), we expect a positive SFG amplitude when the B-N arm-chair crystal direction points along the positive *x* axis (Fig. 3a,b, φ = 0). This suggests that the vast majority of hBN islands have N-terminated zigzag edges, Fig. 3e,f(i), whereas we do not observe B-terminated edges, Fig. 3f(iii). While this was previously concluded from local measurements with STM, TEM, and calculations for triangular-shaped islands,[38-40] phase-resolved SFG microscopy with azimuthal scanning enables us to measure the edge termination across much larger length scales, as well as a simultaneous characterization of many hBN islands.

## Discussion

Overall, we have shown that phase-resolved SFG microscopy in combination with azimuthal scanning enables full crystallographic imaging of hBN monolayers with IR-subdiffractional spatial resolution. Compared to SHG, SFG is strongly enhanced by matching the IR laser wavelength with a phonon resonance of the material that is both IR and Raman-active. This makes SFG of hBN monolayers equally efficient as SHG of excitonic 2D materials and thereby enables rapid screening of this usually invisible layer, with sensitivity to crystal orientation and more generally also layer number and stacking configuration. The technique further naturally provides images across a range of IR frequencies that enable chemical selectivity and allow to distinguish different materials. SFG microscopy may thus also be used to image strain distributions,[41] with combined sensitivity to magnitude and direction by probing phonon frequency shifts and polarization selection rules, as well as to image frequency-dependent propagation of polaritons.[34,42] As a far-field optical technique that is selective to hBN through its IR phonon resonance, SFG microscopy has the potential to image spatial heterogeneity and twist angles of hBN inside of van der Waals heterostructures and devices in a non-invasive and label-free way. More generally, the technique is also applicable to a wide range of other van der Waals materials with broken inversion symmetry, interfaces between materials, as well as combinations with anisotropic molecular assemblies.

Furthermore, the large phonon-enhanced second-order nonlinearity of monolayer hBN is particularly appealing for converting mid-IR to visible light, which could enable new optoelectronic devices and detectors based on vdW heterostructures. The conversion efficiency could be even further increased by using 3R-stacked hBN,[43] where the nonlinear intensity is expected to increase quadratically with layer number,[44] and by nanophotonic structures,[34,45] potentially enabling continuous-wave upconversion or visible detection of thermal radiation.[46,47] The sensitivity to stacking order makes SFG microscopy also promising for rapid screening and in-situ imaging of sliding ferroelectricity, where the inversion symmetry of hBN bi- and multilayers is broken through their local stacking reconfiguration.[43,48,49] The nonlinear response of hBN can be further tailored through nonlinear phonon-mediated processes that are expected at large IR laser intensities, such as polariton blockade in nanometer-sized structures and strong frequency shifts by mixing static and optical fields.[35] Such phonon-driven effects, also including helical and chiral nonlinear phononics,[50] could be characterized with pump-probe measurements, which are compatible with our SFG time-domain approach.[51]

# Methods

**Phase-Resolved SFG Microscopy**

The details of the azimuthal-scanning, phase-resolved SFG microscope used in this work can be found elsewhere in previous works.[21,22,25] In short, the ~30 fs, ~800 nm output of a 1 kHz Ti:sapphire laser (~7 W) is split in two, with each beam seeding two independent optical parametric amplifiers (OPA). The signal and idler of the first OPA are then combined in a nonlinear crystal to produce their tunable difference-frequency (DFG) output in the mid-IR, and the signal of the second OPA is frequency-doubled to produce a 690 nm (VIS) beam. The mid-IR beam is then split by a 45° (s-polarized) KBr beam splitter, with the weaker part being combined collinearly with the VIS and directed through a *z*-cut α-quartz window for LO generation. The transmitted VIS and output LO beams then pass by a delay stage and are subsequently combined collinearly with the stronger part of the mid-IR and sent to the microscope.

In the microscope setup, a single collinear beam comprising mid-IR (p-polarized), VIS (p-polarized), and LO (s-polarized) pulses is focused onto the sample at an incidence angle of 36° relative to the surface normal using a parabolic mirror. The beams are directed through a custom-made hole in a reflective Schwarzschild objective (40x, PIKE Technologies, 0.78 NA). The SFG and reflected light is collected by the objective, filtered to remove all frequencies except the SFG (and LO) and focused onto a thermoelectrically cooled EMCCD camera. To implement balanced imaging,[21] before the camera, the SFG and LO polarizations are rotated by 45° and then split by a polarizing beam splitter to produce two images with opposing interference ($\pm$), with intensity

$$I_{\text{het}}(\Delta t) \propto |E_{\text{SFG}}(\Delta t)|^2 + |E_{\text{LO}}|^2 \pm 2 E_{\text{SFG}}(\Delta t) E_{\text{LO}}, \qquad (2)$$

with electric fields $E_{\text{SFG}}$ and $E_{\text{LO}}$ of the SFG and LO. Both images are recorded simultaneously on the same CCD array and their difference is calculated (using a paired pixel calibration measurement) to isolate the cross-term response $S_{\text{SFG}} \propto E_{\text{SFG}} E_{\text{LO}}$.

The SFG images and spectra presented in this work were acquired interferometrically with 5 fs time steps from a mid-IR – VIS delay time from -300 to 2500 fs. Each image was acquired for 1.5 s (unless otherwise noted) and averaged over 3 exposures. The full interferometric trace was recorded for azimuthal sample rotations in 15° steps ranging from 0 to 345°.

**Data analysis**

The obtained interferometric SFG images were firstly treated to remove any non-zero offset in the interferograms, scaled by a Hanning window, and zero-padded to double the number of time points.

The result was then converted into spectral frequencies via a Fourier transformation and normalized in both amplitude and phase by a reference measurement of *z*-cut α-quartz (5 fs steps, -300 to 300 fs range, 1.5 s exposures, 3 averages). A singular value decomposition of the normalized data was then used to remove high spatial frequency imaging artifacts arising from the microscope geometry. These steps were implemented in the software Igor Pro.

The treated hyperspectral images for each sample rotation were then back-rotated by the known rotation angle and spatially matched using the cross-correlation of the simultaneously obtained linear reflection images of the LO, having removed the illumination envelope via 2D Fourier filtering. A complementary rotational matching approach based on machine learning produced similar results. The overlapped images were then subjected to a further rotational Fourier transformation to convert the azimuthal rotation angles into azimuthal frequencies,[25] isolating the three-fold frequency component that contains the crystallographic information of the monolayer hBN which has $D_{3h}$ symmetry (see SI Section S3).

**Growth and Transfer of hBN Monolayers**

hBN was grown via low-pressure chemical vapor deposition (CVD) in a custom-built hot-walled system[24,28] on an iron (Fe) foil catalyst. Prior to growth, the Fe foil was etched in 0.2 M $FeCl_3$ for 5 minutes to remove surface contaminants, then rinsed in water and dried before loading into the reactor. The Fe foil was first annealed at a temperature of 1050 °C for 30 minutes under 50 sccm of $H_2$ (~0.5 Torr) to reduce iron oxides and enlarge Fe grains. Under 50 sccm of $H_2$ at 1050 °C, the solid precursor (ammonia-borane, ~3 mg) was heated to 90 °C upstream of the reactor, introducing B and N-containing vapor species into the reactor to initiate hBN growth. After 30 minutes of growth, the precursor supply was shut off, and the system was cooled under $H_2$. Figure S7 shows a scanning electron microscopy image of hBN islands on the Fe catalyst.

hBN was transferred to fused silica (500 μm thick) via a sacrificial PMMA layer. A PMMA solution (4 wt% in anisole) was first spin coated onto the hBN/Fe surface and allowed to dry. The Fe was then etched overnight by floating on 0.2 M $FeCl_3$ solution. After the Fe was completely dissolved, the PMMA/hBN was rinsed with 0.1 M HCl to remove $FeCl_3$ and iron-oxide residues before rising three times with DI water. The PMMA/hBN was then scooped onto the target fused silica, dried, then heated to 90 °C to remove remaining water residues and ensure good adhesion. The PMMA layer was then dissolved overnight in acetone, the sample rinsed in isopropyl alcohol and dried.

**AFM Imaging**

The surface topography was characterised using an atomic force microscope (AFM, XE-150 from Park Systems) in non-contact mode with a scan rate of 0.23 Hz. We used a silicon tip (PPP-NCHR) with 330 kHz tapping frequency and a set point of 12.7 nm. The images presented here are a cut-out of a 45x45 µm scan with 1024 lines.

# Acknowledgements

P.R.K. acknowledges support from DOE Early Career Research Program award #DE-SC002291 and NSF CAREER award #1944134. R.A.K acknowledges support by the NASA Space Technology Graduate Research Opportunity and J.D.C support by the Office of Naval Research MURI on Twist-Optics under grant N0001-23-1-2567. D.B., H.H.H., and K.R. acknowledge the European Union's Horizon 2020 research and innovation program under Grant Agreement No. 101135168 (2D-ENGINE).

# Author Contributions

The project was conceived by A.P., M.T. and M.W. The SFG measurements were conducted by A.P.F., N.S.M. and B.J., and data were analyzed by A.P.F., N.S.M. and D.B. The samples were prepared by A.E.N., P.R.K., R.A.K. and J.D.C. AFM measurements were conducted by K.G.H. and N.S.M. DFT calculations were conducted by C.C. The data analysis was supported by D.B., H.H.H., C.S. and K.R. The manuscript was written by N.S.M. with contributions from all co-authors.

# References

1   Dean, C. R. *et al.* Boron nitride substrates for high-quality graphene electronics. *Nature Nanotechnology* **5**, 722-726, doi:10.1038/nnano.2010.172 (2010).
2   Roy, S. *et al.* Structure, Properties and Applications of Two-Dimensional Hexagonal Boron Nitride. *Advanced Materials* **33**, 2101589, doi:10.1002/adma.202101589 (2021).
3   Caldwell, J. D. *et al.* Photonics with hexagonal boron nitride. *Nature Reviews Materials* **4**, 552-567, doi:10.1038/s41578-019-0124-1 (2019).
4   Tran, T. T., Bray, K., Ford, M. J., Toth, M. & Aharonovich, I. Quantum emission from hexagonal boron nitride monolayers. *Nature Nanotechnology* **11**, 37-41, doi:10.1038/nnano.2015.242 (2016).
5   Caldwell, J. D. *et al.* Sub-diffractional volume-confined polaritons in the natural hyperbolic material hexagonal boron nitride. *Nature Communications* **5**, 5221, doi:10.1038/ncomms6221 (2014).
6   Dai, S. *et al.* Tunable Phonon Polaritons in Atomically Thin van der Waals Crystals of Boron Nitride. *Science* **343**, 1125-1129, doi:10.1126/science.1246833 (2014).
7   Li, P. *et al.* Hyperbolic phonon-polaritons in boron nitride for near-field optical imaging and focusing. *Nature Communications* **6**, 7507, doi:10.1038/ncomms8507 (2015).
8   Gorbachev, R. V. *et al.* Hunting for Monolayer Boron Nitride: Optical and Raman Signatures. *Small* **7**, 465-468, doi:10.1002/smll.201001628 (2011).
9   Ling, J. *et al.* Vibrational Imaging and Quantification of Two-Dimensional Hexagonal Boron Nitride with Stimulated Raman Scattering. *ACS Nano* **13**, 14033-14040, doi:10.1021/acsnano.9b06337 (2019).

10. Lin, E. *et al.* Hyperspectral microscopy of boron nitride nanolayers using hybrid femto/picosecond coherent anti-Stokes Raman scattering. *Opt. Lett.* **49**, 2329-2332, doi:10.1364/OL.519571 (2024).
11. Li, Y. *et al.* Probing Symmetry Properties of Few-Layer MoS2 and h-BN by Optical Second-Harmonic Generation. *Nano Letters* **13**, 3329-3333, doi:10.1021/nl401561r (2013).
12. Kim, S. *et al.* Second-harmonic generation in multilayer hexagonal boron nitride flakes. *Opt. Lett.* **44**, 5792-5795, doi:10.1364/OL.44.005792 (2019).
13. Yao, K. *et al.* Enhanced tunable second harmonic generation from twistable interfaces and vertical superlattices in boron nitride homostructures. *Science Advances* **7**, eabe8691, doi:10.1126/sciadv.abe8691 (2021).
14. Zhang, T. *et al.* Accurate Layer-Number Determination of Hexagonal Boron Nitride Using Optical Characterization. *Nano Letters* **24**, 14774-14780, doi:10.1021/acs.nanolett.4c04241 (2024).
15. Zhou, L. *et al.* Nonlinear Optical Characterization of 2D Materials. *Nanomaterials* **10**, 2263, doi:10.3390/nano10112263 (2020).
16. Yin, X. *et al.* Edge Nonlinear Optics on a MoS2 Atomic Monolayer. *Science* **344**, 488-490, doi:10.1126/science.1250564 (2014).
17. Kim, W., Ahn, J. Y., Oh, J., Shim, J. H. & Ryu, S. Second-Harmonic Young's Interference in Atom-Thin Heterocrystals. *Nano Letters* **20**, 8825-8831, doi:10.1021/acs.nanolett.0c03763 (2020).
18. Huang, W., Xiao, Y., Xia, F., Chen, X. & Zhai, T. Second Harmonic Generation Control in 2D Layered Materials: Status and Outlook. *Advanced Functional Materials* **34**, 2310726, doi:10.1002/adfm.202310726 (2024).
19. Xie, Z., Zhao, T., Yu, X. & Wang, J. Nonlinear Optical Properties of 2D Materials and their Applications. *Small* **20**, 2311621, doi:10.1002/smll.202311621 (2024).
20. Zimmermann, J. E., Kim, Y. D., Hone, J. C., Höfer, U. & Mette, G. Directional ultrafast charge transfer in a WSe2/MoSe2 heterostructure selectively probed by time-resolved SHG imaging microscopy. *Nanoscale Horizons* **5**, 1603-1609, doi:10.1039/D0NH00396D (2020).
21. Khan, T. *et al.* Compact oblique-incidence nonlinear widefield microscopy with paired-pixel balanced imaging. *Opt. Express* **31**, 28792-28804, doi:10.1364/OE.495903 (2023).
22. Fellows, A. P., John, B., Wolf, M. & Thämer, M. Spiral packing and chiral selectivity in model membranes probed by phase-resolved sum-frequency generation microscopy. *Nature Communications* **15**, 3161, doi:10.1038/s41467-024-47573-1 (2024).
23. Naclerio, A. E. & Kidambi, P. R. A Review of Scalable Hexagonal Boron Nitride (h-BN) Synthesis for Present and Future Applications. *Advanced Materials* **35**, 2207374, doi:10.1002/adma.202207374 (2023).
24. Naclerio, A. E. *et al.* Scalable Bottom-Up Synthesis of Nanoporous Hexagonal Boron Nitride (h-BN) for Large-Area Atomically Thin Ceramic Membranes. *Nano Letters* **25**, 3221-3232, doi:10.1021/acs.nanolett.4c05939 (2025).
25. Fellows, A. P., John, B., Wolf, M. & Thämer, M. Extracting the Heterogeneous 3D Structure of Molecular Films Using Higher Dimensional SFG Microscopy. *The Journal of Physical Chemistry Letters* **15**, 10849-10857, doi:10.1021/acs.jpclett.4c02679 (2024).
26. Thämer, M., Campen, R. K. & Wolf, M. Detecting weak signals from interfaces by high accuracy phase-resolved SFG spectroscopy. *Physical Chemistry Chemical Physics* **20**, 25875-25882, doi:10.1039/C8CP04239J (2018).
27. Wang, H., Gao, T. & Xiong, W. Self-Phase-Stabilized Heterodyne Vibrational Sum Frequency Generation Microscopy. *ACS Photonics* **4**, 1839-1845, doi:10.1021/acsphotonics.7b00411 (2017).
28. Kidambi, P. R. *et al.* In Situ Observations during Chemical Vapor Deposition of Hexagonal Boron Nitride on Polycrystalline Copper. *Chemistry of Materials* **26**, 6380-6392, doi:10.1021/cm502603n (2014).
29. Niemann, R. *et al.* Long-wave infrared super-resolution wide-field microscopy using sum-frequency generation. *Applied Physics Letters* **120**, 131102, doi:10.1063/5.0081817 (2022).
30. Thämer, M., Garling, T., Campen, R. K. & Wolf, M. Quantitative determination of the nonlinear bulk and surface response from alpha-quartz using phase sensitive SFG spectroscopy. *The Journal of Chemical Physics* **151**, 064707, doi:10.1063/1.5109868 (2019).

31  Ginsberg, J. S. *et al.* Phonon-enhanced nonlinearities in hexagonal boron nitride. *Nature Communications* **14**, 7685, doi:10.1038/s41467-023-43501-x (2023).
32  Liu, W.-T. & Shen, Y. R. Sum-frequency phonon spectroscopy on alpha-quartz. *Physical Review B* **78**, 024302, doi:10.1103/PhysRevB.78.024302 (2008).
33  Roman, E., Yates, J. R., Veithen, M., Vanderbilt, D. & Souza, I. Ab initio study of the nonlinear optics of III-V semiconductors in the terahertz regime. *Physical Review B* **74**, 245204, doi:10.1103/PhysRevB.74.245204 (2006).
34  Niemann, R. *et al.* Spectroscopic and Interferometric Sum-Frequency Imaging of Strongly Coupled Phonon Polaritons in SiC Metasurfaces. *Advanced Materials* **36**, 2312507, doi:10.1002/adma.202312507 (2024).
35  Iyikanat, F., Konečná, A. & García de Abajo, F. J. Nonlinear Tunable Vibrational Response in Hexagonal Boron Nitride. *ACS Nano* **15**, 13415-13426, doi:10.1021/acsnano.1c03775 (2021).
36  Kim, W. *et al.* Exciton-Sensitized Second-Harmonic Generation in 2D Heterostructures. *ACS Nano* **17**, 20580-20588, doi:10.1021/acsnano.3c07428 (2023).
37  Kim, K. K. *et al.* Synthesis of Monolayer Hexagonal Boron Nitride on Cu Foil Using Chemical Vapor Deposition. *Nano Letters* **12**, 161-166, doi:10.1021/nl203249a (2012).
38  Liu, Y., Bhowmick, S. & Yakobson, B. I. BN White Graphene with "Colorful" Edges: The Energies and Morphology. *Nano Letters* **11**, 3113-3116, doi:10.1021/nl2011142 (2011).
39  Auwärter, W., Suter, H. U., Sachdev, H. & Greber, T. Synthesis of One Monolayer of Hexagonal Boron Nitride on Ni(111) from B-Trichloroborazine (ClBNH)3. *Chemistry of Materials* **16**, 343-345, doi:10.1021/cm034805s (2004).
40  Ryu, G. H. *et al.* Atomic-scale dynamics of triangular hole growth in monolayer hexagonal boron nitride under electron irradiation. *Nanoscale* **7**, 10600-10605, doi:10.1039/C5NR01473E (2015).
41  Mennel, L. *et al.* Optical imaging of strain in two-dimensional crystals. *Nature Communications* **9**, 516, doi:10.1038/s41467-018-02830-y (2018).
42  Frischwasser, K. *et al.* Real-time sub-wavelength imaging of surface waves with nonlinear near-field optical microscopy. *Nature Photonics* **15**, 442-448, doi:10.1038/s41566-021-00782-2 (2021).
43  Wang, L. *et al.* Bevel-edge epitaxy of ferroelectric rhombohedral boron nitride single crystal. *Nature* **629**, 74-79, doi:10.1038/s41586-024-07286-3 (2024).
44  Zhao, M. *et al.* Atomically phase-matched second-harmonic generation in a 2D crystal. *Light: Science & Applications* **5**, e16131-e16131, doi:10.1038/lsa.2016.131 (2016).
45  Zograf, G. *et al.* Combining ultrahigh index with exceptional nonlinearity in resonant transition metal dichalcogenide nanodisks. *Nature Photonics* **18**, 751-757, doi:10.1038/s41566-024-01444-9 (2024).
46  Trovatello, C. *et al.* Quasi-phase-matched up- and down-conversion in periodically poled layered semiconductors. *Nature Photonics* **19**, 291-299, doi:10.1038/s41566-024-01602-z (2025).
47  Ma, R., Yan, H., Zhou, Z., Yu, Y. & Wan, W. Nonlinear upconverted thermal emission through difference frequency generation. *Opt. Lett.* **49**, 4565-4568, doi:10.1364/OL.529620 (2024).
48  Yasuda, K., Wang, X., Watanabe, K., Taniguchi, T. & Jarillo-Herrero, P. Stacking-engineered ferroelectricity in bilayer boron nitride. *Science* **372**, 1458-1462, doi:10.1126/science.abd3230 (2021).
49  Vizner Stern, M. *et al.* Interfacial ferroelectricity by van der Waals sliding. *Science* **372**, 1462-1466, doi:10.1126/science.abe8177 (2021).
50  Minakova, O. *et al.* Direct observation of angular momentum transfer among crystal lattice modes. *arXiv* **2503.11626**, doi:10.48550/arXiv.2503.11626 (2025).
51  Bonn, M. *et al.* Femtosecond Surface Vibrational Spectroscopy of CO Adsorbed on Ru(001) during Desorption. *Physical Review Letters* **84**, 4653-4656, doi:10.1103/PhysRevLett.84.4653 (2000).